\documentclass[a4paper,12pt]{article}
\usepackage{amssymb}
\usepackage{graphicx}
\usepackage{epsfig}
\usepackage{array}
\usepackage{booktabs}
\usepackage{multirow}
\begin{document}
\title{Momentum Fractions of Quarks and Gluons in a Self-similarity based model of Proton}
\author{ $ \mathrm{Akbari \; Jahan}^{\star} $ and D K Choudhury\\ Department of Physics, Gauhati University,\\ Guwahati - 781 014, Assam, India.\\ $ {}^{\star} $Email: akbari.jahan@gmail.com}
\date{}
\maketitle
\begin{abstract}
In this paper, we use momentum sum rule to compute the fractions of momentum carried by quarks and gluons in a self-similarity based model of proton. Comparison of the results with the prediction of QCD asymptotics is also made.\\

\textbf{Keywords \,:} Self-similarity, quarks, gluons.\\

\textbf{PACS Nos.: 05.45.Df ; 24.85.+p}
\end{abstract}
\section{Introduction}
Self-similarity is a possible feature of multi-partons inside a proton at small \textit{x} suggested by Lastovicka \cite{1} and pursued by us later \cite{2,3}. We use momentum sum rule \cite{4} to compute the fractions of momentum carried by quarks and gluons in such a model. As this sum rule needs information about \textit{x} in the regime $ 0 \leq x \leq 1 $, we extrapolate the validity of the model beyond its original range $ 6.2 \, \mathrm{X} \, 10^{-7} \leq x \leq 0.01 $. In Section 2, we outline the formalism and in Section 3, we discuss the results.

\section{Formalism}
\subsection{Self-similarity based model at small \textit{x}}
The model is described by self-similarity based parton density functions (PDFs) in Ref. [1]:
\begin{equation}
q_{i}\left( x,Q^{2}\right)=\frac{e^{D_{0}^{i}}\, Q_{0}^{2}\, x^{-D_{2}}}{1+D_{3}+D_{1}\log \left(\frac{1}{x}\right)}\left(\left( \frac{1}{x} \right)^{D_{1} \log \left( 1+ \frac{Q^{2}}{Q_{0}^{2}}\right) } \left( 1+ \frac{Q^{2}}{Q_{0}^{2}}\right)^{D_{3}+1} -1 \right) 
\end{equation}
where the parameters are
\begin{eqnarray}
D_{1} & = & 0.073\pm 0.001 \nonumber \\
D_{2} & = & 1.013\pm 0.01 \nonumber \\
D_{3} & = & -1.287\pm 0.01 \nonumber \\
Q_{0}^{2} & = & 0.062\pm 0.01 \; \mathrm{GeV}^{2}
\end{eqnarray}
i.e.
\begin{equation}
q_{i}\left( x,Q^{2}\right) =e^{D_{0}^{i}} f\left(x,Q^{2} \right)
\end{equation}
where $ D_{0}^{i} $ is the only flavor dependent parameter and
\begin{equation}
f\left( x,Q^{2}\right)=\frac{ Q_{0}^{2} \, x^{-D_{2}}}{1+D_{3}+D_{1}\log \left(\frac{1}{x}\right)}\left(\left( \frac{1}{x} \right)^{D_{1} \log \left( 1+ \frac{Q^{2}}{Q_{0}^{2}}\right) } \left( 1+ \frac{Q^{2}}{Q_{0}^{2}}\right)^{D_{3}+1} -1 \right) 
\end{equation}
is the flavor independent function for describing \textit{x} and $ Q^{2} $ dependence of the PDF.\\
Defining structure function
\begin{equation}
F_{2}\left( x,Q^{2}\right) = x\sum_{i}e_{i}^{2}\left( q_{i}\left( x,Q^{2}\right) +\bar{q_{i}}\left( x,Q^{2}\right) \right) 
\end{equation}
we get
\begin{equation}
F_{2}\left( x,Q^{2}\right)=e^{D_{0}} \left[ \frac{ Q_{0}^{2}\; x^{-D_{2}+1}}{1+D_{3}+D_{1}\log \left(\frac{1}{x}\right)}\left(\left( \frac{1}{x} \right)^{D_{1} \log \left( 1+ \frac{Q^{2}}{Q_{0}^{2}}\right) } \left( 1+ \frac{Q^{2}}{Q_{0}^{2}}\right)^{D_{3}+1} -1 \right) \right] 
\end{equation}
i.e.
\begin{equation}
F_{2}\left( x,Q^{2}\right) =e^{D_{0}}\, x \, f\left( x,Q^{2} \right)
\end{equation}
where
\begin{equation}
e^{D_{0}}=\sum_{i=1}^{n_{f}} e_{i}^{2}\left( e^{D_{0}^{i}}+ e^{\bar{D_{0}^{i}}}\right) 
\end{equation}
with $ D_{0} = 0.339 \pm 0.145 $.

\subsection{Momentum sum rule}
The momentum sum rule \cite{4} is given as
\begin{equation}
\int\limits_{0}^{1}x \sum\left( q_{i}\left( x,Q^{2}\right)+ \bar{q_{i}}\left( x,Q^{2}\right) \right)\, dx + \int\limits_{0}^{1}G\left(x,Q^{2}\right)\, dx =1
\end{equation}
where
\begin{equation}
\langle x \rangle_{q}=\int\limits_{0}^{1} \sum\left( q_{i}\left( x,Q^{2}\right)+ \bar{q_{i}}\left( x,Q^{2}\right) \right)\, dx
\end{equation}
and
\begin{equation}
\langle x \rangle_{g}=\int\limits_{0}^{1} G\left(x,Q^{2}\right)\, dx
\end{equation}
are the momentum fractions carried by quarks and gluons respectively.\\
Using Eq (3) in Eq (10),
\begin{equation}
\langle x \rangle_{q}= e^{\tilde{D_{0}}} \int\limits_{0}^{1} x \,f\left(x,Q^{2} \right) \, dx
\end{equation}
where
\begin{equation}
e^{\tilde{D_{0}}}=\sum_{i=1}^{n_{f}} \left( e^{D_{0}^{i}}+ e^{\bar{D_{0}^{i}}}\right) 
\end{equation}
Note that $ D_{0} $ and $ \tilde{D_{0}} $ are not identical except in specific model \cite{5, 6,7}, where partons have integral charges. Hence the momentum sum rule as given in Eq (9) becomes
\begin{equation}
\int\limits_{0}^{1}\left(a \, F_{2}\left( x,Q^{2}\right) + G\left( x,Q^{2}\right) \right) \, dx = 1
\end{equation}
where
\begin{equation}
a=\frac{e^{\tilde{D_{0}}}}{e^{D_{0}}}
\end{equation}
which has to be determined from data.\\\\
Since $ D_{3} $ is negative and $ D_{1}$ is positive (Eq (2)), there exists a singularity in \textit{x} at $ x\approx 0.019 $ in the structure function $ F_{2}\left( x,Q^{2}\right)$ (Eq (6)). Any viable structure function is not expected to have such a singularity in the region $ 0 \leq x \leq 1 $. We therefore evaluate the integral of Eq (14) by Cauchy's principle value integration \cite{8} leading to
\begin{equation}
\langle x \rangle_{q}=a.I\left( x_{min},Q^{2} \right) 
\end{equation}
where $ I\left( x_{min},Q^{2} \right) $ is an infinite series.
\begin{eqnarray}
\lefteqn{I\left( x_{min},Q^{2} \right)\equiv \frac{e^{D_{0}}\, Q_{0}^{2}}{D_{1}} \left( 1+ \frac{Q^{2}}{Q_{0}^{2}} \right)^{-D_{3}+1} e^{-\mu y_{0}}. } \nonumber \\
& & \left(\log| \frac{y_{max}-y_{0}}{y_{0}}| -\mu y_{max}+ \sum_{n=2}^{\infty} \frac{(-1)^{n}\mu ^{n}}{n.n!} \left\lbrace \left( y_{max}-y_{0} \right)^{n}+ (-1)^{n+1}y_{0}^{n} \right\rbrace \right) \nonumber \\
& & -\frac{e^{D_{0}}\, Q_{0}^{2}}{D_{1}}.e^{-\rho y_{0}}. \nonumber \\
& & \left(\log\vert \frac{y_{max}-y_{0}}{y_{0}}\vert -\rho y_{max}+ \sum_{n=2}^{\infty} \frac{(-1)^{n}\rho ^{n}}{n.n!} \left\lbrace \left( y_{max}-y_{0} \right)^{n}+ (-1)^{n+1}y_{0}^{n} \right\rbrace \right) \nonumber \\
& & 
\end{eqnarray}
where
\begin{eqnarray}
\alpha & = & D_{2}-1+D_{1}\log \left(1+ \frac{Q^{2}}{Q_{0}^{2}} \right) \nonumber \\
\beta & = & D_{2}-1 \nonumber \\
\mu & = & 1-\alpha = 1-\left( D_{2}-1+ D_{1}\log \left(1+ \frac{Q^{2}}{Q_{0}^{2}}\right) \right) \nonumber \\
\rho & = & 1- \beta = 1 - (D_{2}-1) \nonumber \\
y_{0} & = & \frac{D_{3}+1}{D_{1}} \nonumber \\
y_{max} & = & \log \left( \frac{1}{x_{min}}\right) 
\end{eqnarray}
Here $ x_{min} $ is introduced to take care of the end-point singularity $ x = 0 $ in the model. Taking only the first term of the integral $ I\left( x_{min}, Q^{2} \right) $, we have
\begin{equation}
\langle x \rangle_{q}=a.(0.024607).\log \vert \frac{y_{max}-y_{0}}{y_{0}}\vert. \frac{Q^{2}}{Q_{0}^{2}}
\end{equation}
\section{Results}
In order to make numerical analysis, we have to determine \textit{a} and $ x_{min} $ defined in Eq (15) and Eq (19) respectively. We take
\begin{equation}
\langle x \rangle_{q}=0.465\pm 0.023 \;\;\mathrm{and} \; \int\limits_{0}^{1}F_{2}\left( x,Q^{2} \right) \, dx = 0.148\pm 0.007
\end{equation}
at $ Q^{2}=15 \, \mathrm{GeV^{2}} $ \cite{9} and evaluate
\begin{equation}
a=\frac{\langle x \rangle_{q}}{\int \limits_{0}^{1} F_{2}\left(x,Q^{2}\right)\, dx }\approx 3.1418
\end{equation}
Using it in Eq (19), we obtain
\begin{equation}
x_{min}\approx 3.4953\times 10^{-4}
\end{equation}
In Table 1, we record the values of $\langle x \rangle_{q}$ and $\langle x \rangle_{g}$ for several representative values of $ Q^{2} $. We also show the values with $ a = 1 $ corresponding to partons having integral charges $ \pm 1 $ in order to explicitly show the effects of charges in the partition of momentum among quarks and gluons in the model.
\begin{table}[!h]
\begin{center}
\caption{\textbf{Values of $\langle x\rangle_{q}$ and $\langle x \rangle_{g}$ for a few representative values of $ Q^{2} $.}}
\bigskip
\begin{tabular}{|c|c|c|c|c|}
\hline
\multirow{2}{*}{$ Q^{2}\left(\mathrm{GeV^{2}}\right)$} &  \multicolumn{2}{c}{$\langle x\rangle_{q}$} \vline  & \multicolumn{2}{c}{$\langle x\rangle_{g}$} \vline \\ 
{} & $a=1$   & $a=3.1418$   & $a=1$   & $a=3.1418$ \\
\hline \hline
10 & 0.0716 & 0.22516 & 0.9283 & 0.77484 \\ 
20 & 0.1433 & 0.45032 & 0.8567 & 0.54968 \\ 
30 & 0.2149 & 0.67548 & 0.7850 & 0.32452 \\ 
40 & 0.2867 & 0.90064 & 0.7133 & 0.09936 \\ 
45 & 0.3225 & 1 & 0.6775 & 0 \\ \hline
50 & 0.3583 & \multirow{9}{*}{$ > 1 $} & 0.6418  & \multirow{9}{*}{$ < 0 $} \\ 
60 & 0.4299 & {} & 0.5700 & {} \\ 
65 & 0.4658 & {} & 0.5342 & {} \\ 
70 & 0.5017 & {} & 0.4983 & {} \\ 
75 & 0.5375 & {} & 0.4625 & {} \\ 
80 & 0.5733 & {} & 0.4267 & {} \\  
90 & 0.6449 & {} & 0.3550 & {} \\ \hline
\end{tabular}
\end{center}
\end{table}
\begin{figure}[h]
\centering
\includegraphics[scale=0.5]{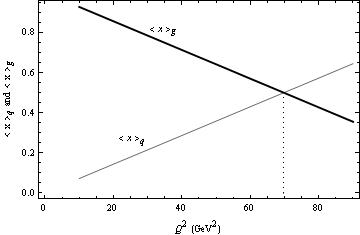}
\caption{$ \langle x \rangle_{q} \, \mathrm{and} \, \langle x \rangle_{g} \, \mathrm{vs} \, Q^{2} \, (for \,\, a = 1).$}
\end{figure}

\begin{figure}[h]
\centering
\includegraphics[scale=0.5]{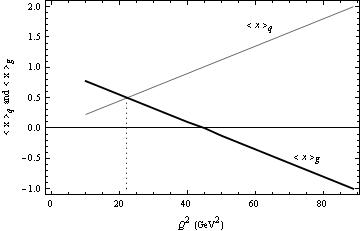}
\caption{$ \langle x \rangle_{q} \, \mathrm{and} \, \langle x \rangle_{g} \, \mathrm{vs} \, Q^{2} \, (for \,\, a = 3.1418). $}
\end{figure}
In Figure 1 and Figure 2, it is shown that within the leading term approximation used, $\langle x \rangle_{q} $ increases with $ Q^{2} $ while $\langle x \rangle_{g} $ decreases. For $ a=1 $ (Figure 1), $ \langle x \rangle_{q}\approx \langle x \rangle_{g}$ at $ Q^{2} \approx 70 \, \mathrm{GeV^{2}}$ and for $ a = 3.1418 $ (Figure 2), $ \langle x \rangle_{q}\approx \langle x \rangle_{g}$  at $ Q^{2} \approx 22 \, \mathrm{GeV^{2}}$ i.e. both quarks and gluons share momentum equally. For $ a = 3.1418 $, the pattern is the same but the share of momentum for quarks increases threefold. It is to be noted that one usually expects the other pattern, i.e. $ \langle x \rangle_{g}=\int\limits_{0}^{1} G\left( x,Q^{2}\right) \, dx $ should increase with $ Q^{2} $ and $ \langle x \rangle_{q}=\int\limits_{0}^{1} F_{2}\left( x,Q^{2}\right) \, dx $ should decrease \cite{10}. This feature is presumably due to the overestimation of the large \textit{x} quarks in Eq (6) and the crude approximation of taking only the first term of the infinite series as in Eq (19). Further, as $ Q^{2}$ increases, $ \langle x \rangle_{q}\approx 1 $ at the saturation scale of $ Q^{2}\approx 140 \, \mathrm{GeV^{2}} $ (for a = 1) and $ Q^{2}\approx 45 \, \mathrm{GeV^{2}} $ (for a = 3.1418) respectively, beyond which the momentum sum rule breaks down. Thus the model in the original version has inherent limitation because of the breakdown of the momentum sum rule beyond the regime $ 45\leq Q^{2}\leq 140 \, \mathrm{GeV^{2}}$. For comparison, we record the experimental values of $\langle x \rangle_{q}$ and $\langle x \rangle_{g}$ in Table 2 available in literature.
\begin{table}[!h]
\begin{center}
\caption{}
\bigskip
\begin{tabular}{|c|c|c|c|}
\hline
{\textbf{$ \langle x \rangle_{q} \, / \, \langle x \rangle_{g}$}} & {\textbf{Value}} & {\textbf{$Q^{2} \, \left(\mathrm{GeV^{2}} \right)$}} & {\textbf{Reference}} \\ \hline
$ \langle x \rangle_{q}$ & $0.3\pm 0.008$ & 3.5 & [10] \\ \hline
$ \langle x \rangle_{q}$ & $0.465 \pm 0.023$ & 15 & [9] \\ \hline
$ \langle x \rangle_{q} \, = \, \langle x \rangle_{g}$ & 0.50 & 10 - 20 & [11] \\ \hline
$ \langle x \rangle_{g}$ & $0.43 \pm 0.03$	& 30 - 100 & [12] \\ \hline
\end{tabular}
\end{center}
\end{table}
\\
We also compare the pattern of change of $ \langle x \rangle_{q}$ and $ \langle x \rangle_{g}$ of the present model with standard QCD. Asymptotically $ \left( Q^{2}\rightarrow \infty \right) $, perturbative QCD predictions of $ \langle x \rangle_{q}$ and $ \langle x \rangle_{g}$ \cite{13, 14} are
\begin{equation}
\lim_{Q^{2}\rightarrow \infty} \langle x \rangle_{q}=\frac{3n_{f}}{2n_{g}+3n_{f}}
\end{equation}
\begin{equation}
\lim_{Q^{2}\rightarrow \infty} \langle x \rangle_{g}=\frac{2n_{g}}{2n_{g}+3n_{f}}
\end{equation}
where $ n_{f} $ is the number of active flavors and $ n_{g} $ is the number of gluons. For $ \mathrm{SU(3)_{c}} $, $n_{g}=8$.\\
In Table 3, we evaluate $ \langle x \rangle_{q} $ and $ \langle x \rangle_{g} $ for various flavors $ n_{f} = 3, 4, 5, 6 $ setting $ n_{g}=8 $ using Eq (23) and Eq (24).
\begin{table}[!h]
\begin{center}
\caption{\textbf{Flavor dependence of asymptotic prediction of $ \langle x \rangle_{q}$ and $ \langle x \rangle_{g}$ in perturbative QCD.}}
\bigskip
\begin{tabular}{|c|c|c|}
\hline
{$ n_{f} $} & {$ \langle x \rangle_{q} $} & {$ \langle x \rangle_{g} $} \\ \hline
3 & 9/25 & 16/25 \\ \hline
4 & 3/7 & 4/7 \\ \hline
5 & 15/31 & 16/31 \\ \hline
6 & 9/17 & 8/17 \\ \hline
\end{tabular}
\end{center}
\end{table}
\\
This table shows that only for $n_{f}=6, \langle x \rangle_{q} \, > \, \langle x \rangle_{g}$.\\

In the present model, on the other hand, such feature sets in even at much lower $Q^{2}$ range as discussed earlier. The study of flavor dependence of $ \langle x \rangle_{q}$ and $ \langle x \rangle_{g}$ in the present model is not possible as it is absorbed in the model parameter $D_{0}$ (Eq (8)). The effects of large \textit{x} and higher order terms are currently under study.\\

It is also interesting to note that the asymptotic partitions of quark and gluon momenta of Eq (23) and Eq (24) have been questioned in recent literature \cite{15}. The suggested alternative asymptotic partition is
\begin{equation}
\lim_{Q^{2}\rightarrow \infty} \langle x \rangle_{q}=\frac{6n_{f}}{n_{g}+6n_{f}}
\end{equation}
\begin{equation}
\lim_{Q^{2}\rightarrow \infty} \langle x \rangle_{g}=\frac{n_{g}}{n_{g}+6n_{f}}
\end{equation}
leading to $ \langle x \rangle_{g}\approx \displaystyle{\frac{4}{15}} \langle x \rangle_{q}$ and  $ \langle x \rangle_{g}\approx \displaystyle{\frac{2}{9}} \langle x \rangle_{q}$ for $n_{f}=5 $ and $n_{f}=6 $ respectively even asymptotically. In the present model such specific partitions set in at finite $Q^{2}$. Thus the main feature of the self-similarity based present analysis, i.e. $ \langle x \rangle_{q}\, > \, \langle x \rangle_{g}$ with increased $Q^{2}$ might conform to the QCD expectation as well if the result of Ref. [15] is taken seriously.\\

The model, even though predicts such interesting results, has several deficiencies of its own. First, the model has phenomenological validity only in a limited $ x $-range. Second, the model falls short of accommodating momentum sum rule for large $Q^{2}$, $Q^{2} \, > \, 140 \, \mathrm{GeV^{2}}$. Finally, as it is purely self-similarity inspired, the model does not conform to standard DGLAP evolution.

\section{Conclusions}
In this paper, we have used momentum sum rule to compute the fractions of momentum carried by quarks and gluons in a self-similarity based model of proton. The main feature of the present analysis is that the  momentum fraction of quarks increases with $Q^{2}$, while the corresponding fraction of gluons decreases. This prediction conforms to asymptotic QCD expectations, as reported recently \cite{15}. The difference is that such momentum partitions set in at finite $Q^{2}$ in the present model.\\

\end{document}